# Photovoltaic effect and photopolarization in Pb[(Mg$_{1/3}$Nb$_{2/3}$)$_{0.68}$Ti$_{0.32}$]O$_3$ crystal

A. S. Makhort,[1] F. Chevrier,[1] D. Kundys,[2] B. Doudin,[1] and B. Kundys[1,*]

[1]*Institut de Physique et Chimie des Matériaux de Strasbourg, UMR 7504 CNRS-UdS, 23 rue du Loess, 67034 Strasbourg, France*
[2]*Scottish Universities Physics Alliance (SUPA), Institute of Photonics and Quantum Sciences, School of Engineering and Physical Sciences, Heriot-Watt University, Edinburgh EH14 4AS, United Kingdom*

Ferroelectric materials are an alternative to semiconductor-based photovoltaics and offer the advantage of above bandgap photovoltage generation. However, there are few known compounds, and photovoltaic efficiencies remain low. Here, we report the discovery of a photovoltaic effect in undoped lead magnesium niobate–lead titanate crystal and a significant improvement in the photovoltaic response under suitable electric fields and temperatures. The photovoltaic effect is maximum near the electric-field-driven ferroelectric dipole reorientation, and increases threefold near the Curie temperature ($T_c$). Moreover, at ferroelectric saturation, the photovoltaic response exhibits clear remanent and transient effects. The transient-remanent combinations together with electric and thermal tuning possibilities indicate photoferroelectric crystals as emerging elements for photovoltaics and optoelectronics, relevant to all-optical information storage and beyond.

Photoferroelectrics are remarkable materials that have great potential for multifunctional applications [1]. These materials that exhibit (multi)ferroic order are particularly interesting, because they offer advanced electric operation that is related to multiple electric polarization switching states [2]. The field was rejuvenated after the discovery of the photovoltaic effect in the multiferroic BiFeO$_3$ [3,4], resulting in the revival of ferroelectric-based photovoltaic operation and related materials [5–8]. Based on recent progress in photovoltaic efficiency of Bi$_2$FeCrO$_6$ films [9], ferroelectric (FE) cells might become competitors for conventional photovoltaics in the near future. In this respect, better insight into photoinduced changes of electrical properties over a wide range of temperatures and electric fields would be helpful. Such a study, however, requires high-quality crystals that are free from the surface/interface effects occurring in thin films [10] and the grain size dependence occurring in ceramics [11]. This task is challenging because the total number of currently known photovoltaic-ferroelectric compounds is well below 20 [5–8]. In this work, we consider a ferroelectric material of the PMN-PT family with the general formula Pb[(Mg$_{1/3}$Nb$_{2/3}$)$_x$Ti$_{1-x}$]O$_3$. Based on an analogy with WO$_3$-doped Pb$_{1-x}$La$_x$(Zr$_y$Ti$_z$)$_{1-x/4}$O$_3$ ceramics [12], the WO$_3$ doping procedure was reported to introduce a photovoltaic effect in Pb(Mg$_{1/3}$Nb$_{2/3}$)$_{0.64}$Ti$_{0.36}$O$_3$ (PMN-36%PT) single crystals [13]. Here, we find that the photovoltaic effect exists even in the widely available undoped Pb[(Mg$_{1/3}$Nb$_{2/3}$)$_x$Ti$_{1-x}$]O$_3$ (PMN-32%PT) composition [14]. We also detail the remanent photopolarization properties originating from the photocarrier generation-distribution mechanism, and demonstrate how the photovoltaic response can be tuned by varying the applied electric field and temperature.

The crystals had (001) orientation, and were square shaped with edges along the [010] and [100] directions [Fig. 1(a)]. Independent energy-dispersive x-ray spectroscopy (EDS) analysis confirmed the PMN-32%PT composition of the crystal. The sample's dimensions were 901 $\mu$m × 272 $\mu$m × 2161 $\mu$m, and both electrodes were formed with silver paste covering the edges in the planes parallel to *yz* [Fig. 1(a)]. This experimental geometry was chosen to minimize light power loss and to avoid possible extrinsic contributions related to light-assisted charge injection from electrodes. The hysteresis loop of polarization versus electric field was taken at room temperature by using a homemade quasistatic FE loop tracer, which has been previously employed to study ferroelectricity in multiferroics [15–17]. The sample was irradiated with a 365 nm (3.4 eV) LED with 30 nm spectral linewidth at 13.67 mW of power, in order to investigate the change in the FE polarization response. The temperature of the sample measured by a thermal camera increases by 1.9 K under illumination and such a temperature change made no noticeable difference to the FE loop.

Figure 1 represents the FE properties of the sample under electric field applied along the [100] direction in darkness, and under illumination along the [001] direction. The single FE loop measured in darkness reveals a classical hysteresis resulting in three polarization states: the spontaneous initial state ("0"), and two saturated states with opposite polarization ("1" and "2"). The ferroelectric loop is very reproducible without any noticeable sign of fatigue (tested at least up to ∼20 cycles) confirming the high quality of the FE crystal. Light irradiation induces a large change in the FE loop. The sample evidently becomes a leakier FE with an apparent increase in both FE polarization and FE coercive field [Fig. 1(b)]. This is confirmed by the corresponding FE current in Fig. 1(c).

The difference between the current under the illumination and the current in darkness [Fig. 1(d)] is the photocurrent in the material, and it reveals at least two important features. The first one is that the photocurrent, and therefore the bulk photovoltaic effect, strongly depends on the FE state. The electric-field-dependent measurements performed here on single crystals, free from additional contributions generally present in thin films and ceramics, evidently reflect the poling history dependence of the pure bulk photovoltaic effect. Secondly, the behavior, clearly illustrated here, reveals that the maximum of the photocurrent is observed at the electric field corresponding to the FE dipole reorientation. Such induced photocurrent should impact the remanent FE state as firstly reported by Iurchuk *et al.* [18]. The time dependence of the related photopolarization measured at the three states (0, 1, and 2) shown in Fig. 2 provides insight into the mechanism explaining how photopolarization properties relate to the ferroelectric state of the material. The sample was initially polarized positively (state "1"), by sweeping the electric field from −0.34 MV/m to +0.34 MV/m and to zero, and the polarization was then monitored during periodic illumination. The reduction of polarization is observed with the remanent effect. This effect can be explained by the generation of free carriers by light. The illumination generates charges that distribute along the previously defined polarization direction [100]. The photocarriers diminish the surface charges, and therefore decrease the internal electric field of the material and the sample polarization. After turning off the light, charge trapping makes this decrease partly persist (remanent effect; Fig. 2), leaving the sample in a slightly reduced polarization state. Subsequent illumination pulses reveal only a transient (reversible) effect as they are of the same energy and the remanence was previously achieved.

---
*Corresponding author: kundysATipcms.unistra.fr

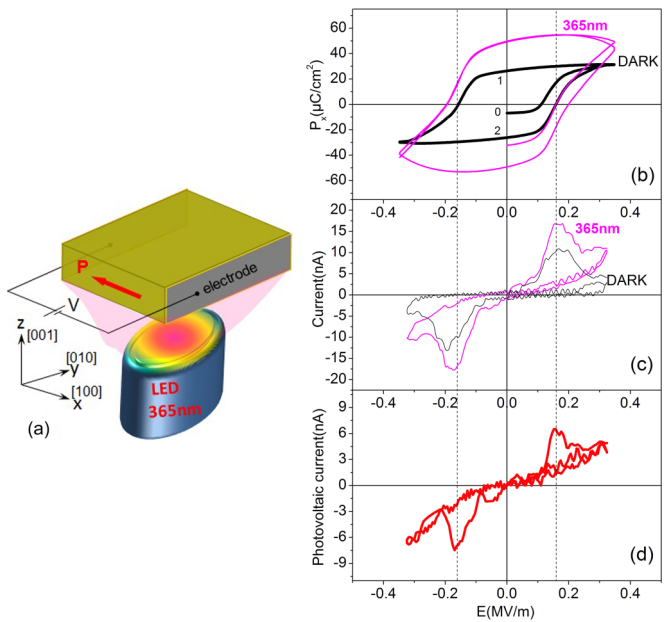

FIG. 1. Photoferroelectric measurements details. (a) Schematic of the experimental setup with respect to the crystalline axis. Ferroelectric polarization (b) and current (c) hysteresis loops in darkness and under 365 nm irradiation. (d) Related photocurrent as a function of electric field.

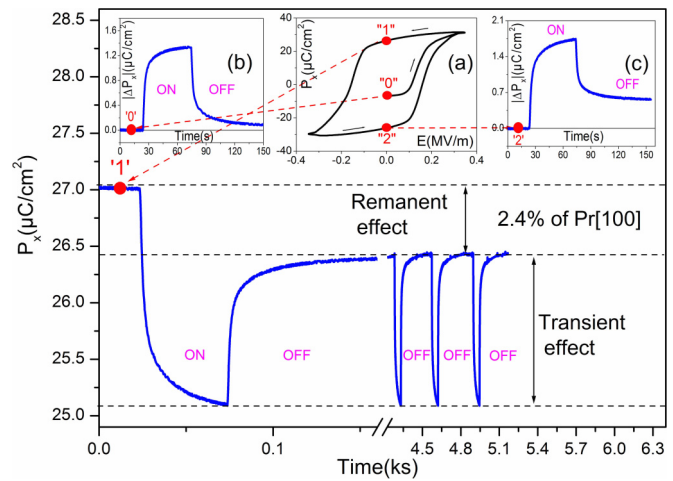

FIG. 2. Photodepolarization effect. A sample initially in remanent state "1," is illuminated by a UV (365 nm) diode. Within ∼50 seconds, a change of polarization is established, with a fraction remaining after turning off the light (remanent effect). Subsequent illumination only reveals the reversible (transient) effect, with no more remanence. Insets show how the remanence in polarization depends on the FE state (a): depolarized (b) and the FE state "2" (c).

Notably, the initial polarization state can be recovered electrically by bringing the ferroelectric system back to the point "1" again in darkness (by sweeping the electric field from −0.34 MV/m to +0.34 MV/m and to zero). As would be expected from the 180° symmetrical polarization rotation, a similar effect, but of opposite sign, is observed for the ferroelectric point "2" (obtained by sweeping the electric field from +0.34 MV/m to −0.34 MV/m and to zero) shown in the inset (c) of Fig. 2. The sample depolarized state "0" obtainable by a damping oscillated voltage procedure [19] shows no remanence [Fig. 2(b)], in support of the photocarriers distribution hypothesis, as in this state an internal electric field is screened by the domain structure. To study the temperature dependence of the photoelectric performance, we have first measured the dielectric and polarization properties in darkness (Fig. 3). Prior to measurements, the sample was first warmed in darkness 19 K above its Curie temperature $T_c$(∼421 K) and then cooled under an applied electric field of 0.237 MV/m. Since the coercive field is much smaller [∼0.16 MV/m; Fig. 1(b)], this procedure ensures a monodomain state by inducing polarization along the [100] direction. The electric field was then set to zero and the dielectric permittivity [Fig. 3(a)] was measured during warming above $T_c$. The same procedure was applied for measuring the electric polarization [Fig. 3(b)].

Three characteristic transitions at $T_1$(∼352 K), $T_2$(∼375 K), and $T_c$(∼ 421 K) are clearly observed. While $T_2$ and $T_c$ are related to the rhombohedral-tetragonal and the tetragonal-cubic transitions [20], respectively, the domain-structure-related anomaly at $T_1$ is known to appear as a function of poling [21]. Taking into account that light can modify the transition temperatures [22–24], the photovoltage was measured for several temperatures, near the above-mentioned critical points where anomalies in the electric properties are observed. The sample was first warmed up to 440 K and then cooled to 300 K under an electric field of 0.27 MV/m to ensure a monodomain FE state. The photovoltaic isotherms were recorded under increasing stabilized temperature values. The results in Fig. 4 show that even at room temperature the transient part of the photovoltage exhibits more than one order of magnitude larger spectral

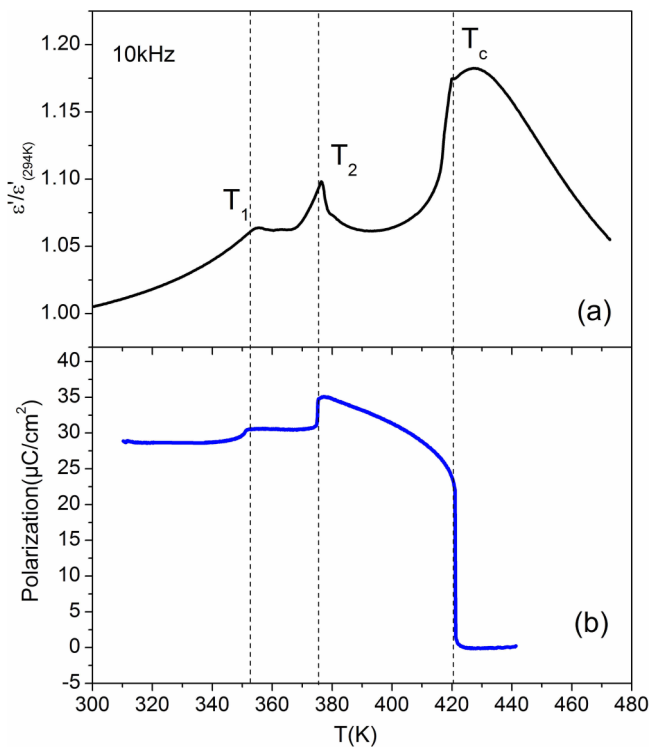

FIG. 3. Temperature dependence of the dielectric permittivity (a) and polarization (b) in darkness.

efficiency than for previously reported $WO_3$-doped crystal [13].

The photovoltaic isotherms reveal nonlinear behavior as a function of light intensity with a characteristic peak. The form of isotherms suggests the occurrence of two competing mechanisms: the light-induced charge generation dominant at low light intensities superposes to the charge recombination processes at higher illumination intensities (Fig. 4). These opposing processes give rise to the peak as a function of light intensity at the value where numbers of generated and recombined carriers become comparable. Because the peak position versus intensity changes with temperature, the maximal photovoltaic change and its linear part have been extracted from the photovoltaic isotherms and plotted against temperature in Fig. 5.

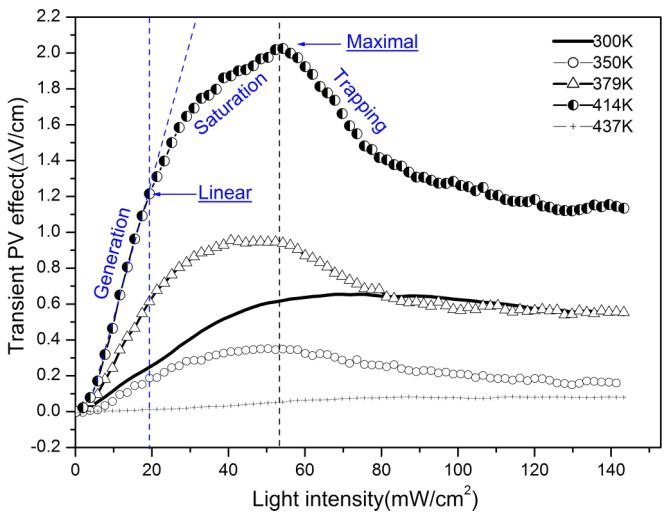

FIG. 4. The transient photovoltage isoterms as a function of light intensity involving charge generation, saturation, and trapping processes.

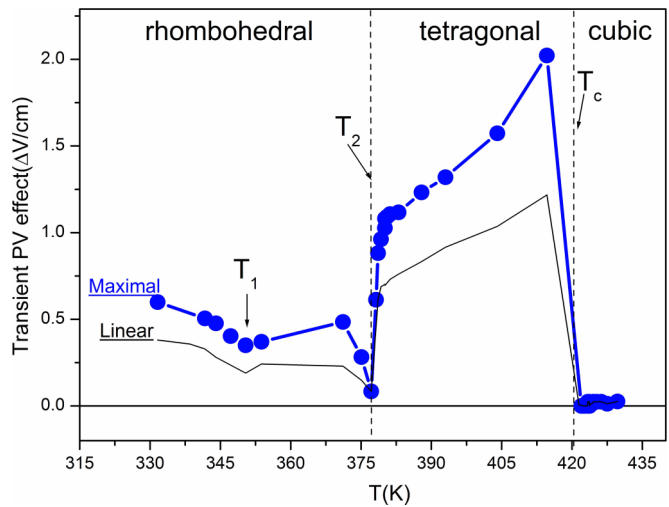

FIG. 5. Temperature dependence of the photovoltaic effect illustrating both linear and maximal parts (see Fig. 4).

The photovoltaic effect shows singularities at all three transition temperatures and reveals more than a threefold enhancement near $T_c$, then vanishing in the paraelectric temperature region. This behavior confirms that photovoltaicity of ferroelectric origin exists in the electrically polar phase only. The generally larger photovoltaic effect in the tetragonal phase may be connected with the expected bandgap temperature dependence reported for similar compounds [25]. The increase of the photovoltaic effect near $T_c$ could, however, be connected to instabilities near the phase transitions, where the FE system is governed by photostrictively [26] correlated fluctuations of the polarization as studied previously [27].

In conclusion, we have discovered that a representative member of the piezoelectric family of PMN-PT complexes, the $Pb[(Mg_{1/3}Nb_{2/3})_{0.68}Ti_{0.32}]O_3$ crystal, exhibits photovoltaic properties in a wide temperature range. The electric field and temperature-controllable photovoltage in this landmark ferroelectric crystal should be regarded as key basic findings, illustrating how charge separation relates to FE properties and superposes on charge recombination. The evident separation between remanent and transient effects that has been found here completes the profile of many recent photoelectric observations [28–33] and is essential to understand and optimize the photoferroelectric performance. This study paves the way toward future investigations in other compounds as well as composites of the same family to optimize composition-to-the-reported properties relationship. In particular, materials used for nanoscale devices [34–36], including magnetic FE compounds [37] with attractive functionalities [38,39] become interesting candidates.

Partial financial support of the Agence Nationale de la Recherche (hvSTRICTSPIN ANR-13-JS 04-207 0008-01, Labex NIE 0058_NIE within the Investissement d'Avenir

Photovoltaic effect and photopolarization in Pb[(Mg1/3Nb2/3)0.68Ti0.32]O3 crystal, Phys. Rev. Materials 2, 012401(R) (2018).

program ANR-10-IDEX-0002-02) is acknowledged. A.M. acknowledges PhD co-fund program of the Alsace region. The authors also very grateful to Jacques Faerber for EDS analysis and Michael Coey for reading the manuscript.
[1] J. Kreisel, M. Alexe, and P. A. Thomas, Nat. Mater. 11, 260 (2012).
[2] D. Lee, S. M. Yang, T. H. Kim, B. C. Jeon, Y. S. Kim, J.-G. Yoon, H. N. Lee, S. H. Baek, C. B. Eom, and T. W. Noh, Adv. Mater. 24, 402 (2012).
[3] T. Choi, S. Lee, Y. J. Choi, V. Kiryukhin, and S.-W. Cheong, Science 324, 63 (2009).
[4] H. T. Yi, T. Choi, S. G. Choi, Y. S. Oh, and S.-W. Cheong, Adv. Mater. 23, 3403 (2011).
[5] K. T. Butler, J. M. Frost, and A. Walsh, Energy Environ. Sci. 8, 838 (2015).
[6] C. Paillard, X. Bai, I. C. Infante, M. Guennou, G. Geneste, M. Alexe, J. Kreisel, and B. Dkhil, Adv. Mater. 28, 5153 (2016).
[7] I. Grinberg, D. V. West, M. Torres, G. Gou, D. M. Stein, L. Wu, G. Chen, E. M. Gallo, A. R. Akbashev, P. K. Davies, J. E. Spanier, and A. M. Rappe, Nature (London) 503, 509 (2013).
[8] P. Lopez-Varo, L. Bertoluzzi, J. Bisquert, M. Alexe, M. Coll, J. Huang, J. A. Jimenez-Tejada, T. Kirchartz, R. Nechache, F. Rosei, and Y. Yuan, Phys. Rep. 653, 1 (2016).
[9] R. Nechache, C. Harnagea, S. Li, L. Cardenas, W. Huang, J. Chakrabartty, and F. Rosei, Nat. Photon. 9, 61 (2015).
[10] M. L. Calzada, R. Jimenez, A. Gonzalez, J. Garcıa-Lopez, D. Leinen, and E. Rodrıguez-Castellon, Chem. Mater. 17, 1441 (2005).
[11] K. Takagi, S. Kikuchi, J.-F. Li, H. Okamura, R. Watanabe, and A. Kawasaki, J. Am. Ceram. Soc. 87, 1477 (2004).
[12] P. Poosanaas, A. Dogan, S. Thakoor, and K. Uchino, J. Appl. Phys. 84, 1508 (1998).
[13] C.-S. Tu, F.-T. Wang, R. R. Chien, V. H. Schmidt, C.-M. Hung, and C.-T. Tseng, Appl. Phys. Lett. 88, 032902 (2006).
[14] PMN-PT crystals are commercially available from www.Crystal-gmbh.com (Germany).
[15] B. Kundys, C. Simon, and C. Martin, Phys. Rev. B 77, 172402 (2008).
[16] B. Kundys, A. Maignan, C. Martin, N. Nguyen, and C. Simon, Appl. Phys. Lett. 92, 112905 (2008).
[17] B. Kundys, A. Lappas, M. Viret, V. Kapustianyk, V. Rudyk, S. Semak, C. Simon, and I. Bakaimi, Phys. Rev. B 81, 224434 (2010).
[18] V. Iurchuk, D. Schick, J. Bran, D. Colson, A. Forget, D. Halley, A. Koc, M. Reinhardt, C. Kwamen, N. A. Morley, M. Bargheer, M. Viret, R. Gumeniuk, G. Schmerber, B. Doudin, and B. Kundys, Phys. Rev. Lett. 117, 107403 (2016).
[19] B. Kundys, V. Iurchuk, C. Meny, H. Majjad, and B. Doudin, Appl. Phys. Lett. 104, 232905 (2014).
[20] T. R. Shrout, Z. P. Chang, N. Kim, and S. Markgraf, Ferroelectr., Lett. Sect. 12, 63 (1990).
[21] Y. Guo, H. Luo, D. Ling, H. Xu, T. He, and Z. Yin, J. Phys.: Condens. Matter 15, L77 (2003).
[22] S. Ueda, I. Tatsuzaki, and Y. Shindo, Phys. Rev. Lett. 18, 453 (1967).
[23] H. Borkar, V. Rao, M. Tomar, V. Gupta, J. F. Scott, and A. Kumar, RSC Adv. 7, 12842 (2017).
[24] H. Borkar, M. Tomar, V. Gupta, R. S. Katiyar, J. F. Scott, and A. Kumar, Mater. Res. Express 4, 086402 (2017).
[25] F. Wu, X. He, J. Zhang, B. Yang, E. Sun, J. Jiang, and W. Cao, Opt. Mater. 60, 101 (2016).
[26] B. Kundys, Appl. Phys. Rev. 2, 011301 (2015).
[27] S. H. Wemple and M. DiDomenico, Phys. Rev. B 1, 193 (1970).
[28] R. Moubah, O. Rousseau, D. Colson, A. Artemenko, M. Maglione, and M. Viret, Adv. Funct. Mater. 22, 4814 (2012).
[29] A. Bhatnagar, Y. H. Kim, D. Hesse, and M. Alexe, Nano Lett. 14, 5224 (2014).
[30] K. Bogle, R. Narwade, A. Phatangare, S. Dahiwale, M. Mahabole, and R. Khairnar, Phys. Status Solidi A 213, 2183 (2016).
[31] Z. Bai, W. Geng, Y. Zhang, S. Xu, H. Guo, and A. Jiang, Appl. Phys. A 123, 561 (2017).
[32] H. Fan, Z. Fan, P. Li, F. Zhang, G. Tian, J. Yao, Z. Li, X. Song, D. Chen, B. Han, M. Zeng, S. Wu, Z. Zhang, M. Qin, X. Lu, J. Gao, Z. Lu, Z. Zhang, J. Dai, X. Gao, and J.-M. Liu, J. Mater. Chem. C 5, 3323 (2017).
[33] B. Kundys, C. Meny, M. R. J. Gibbs, V. Da Costa, M. Viret, M. Acosta, D. Colson, and B. Doudin, Appl. Phys. Lett. 100, 262411 (2012).
[34] S. Y. Yang, J. Seidel, S. J. Byrnes, P. Shafer, C.-H. Yang, M. D. Rossell, P. Yu, Y.-H. Chu, J. F. Scott, J. W. Ager, L. W. Martin, and R. Ramesh, Nat. Nanotechnol. 5, 143 (2010).
[35] W. J. Hu, Z. Wang, W. Yu, and T. Wu, Nat. Commun. 7, 10808 (2016).
[36] J. E. Spanier, V. M. Fridkin, A. M. Rappe, A. R. Akbashev, A. Polemi, Y. Qi, Z. Gu, S. M. Young, C. J. Hawley, D. Imbrenda, G. Xiao, A. L. Bennett-Jackson, and C. L. Johnson, Nat. Photon. 10, 611 (2016).
[37] S. Manz, M. Matsubara, T. Lottermoser, J. Büchi, A. Iyama, T. Kimura, D. Meier, and M. Fiebig, Nat. Photon. 10, 653 (2016).
[38] B. Mettout and P. Gisse, Ferroelectrics 506, 93 (2017).
[39] B. Kundys, M. Viret, D. Colson, and D. O. Kundys, Nat. Mater. 9, 803 (2010).
(4)